\documentclass[twocolumn,prb,showpacs,multicol,amsmath,amssymb]{revtex4}
\usepackage[dvips]{graphicx}
\usepackage{graphicx}
\usepackage{dcolumn}
\usepackage{bm}
\usepackage{graphics}
\usepackage{epsfig,color}

\newcommand{\be}{\begin{equation}}
\newcommand{\ee}{\end{equation}}
\newcommand{\bea}{\begin{eqnarray}}
\newcommand{\eea}{\end{eqnarray}}

\newcommand{\bS}{\bf S}

\begin{document}

\title{Concurrence and  entanglement entropy in a dimerised spin-1/2 two-leg ladder}

\author{Somayyeh Nemati, Saeed Batebi, Saeed Mahdavifar}
\affiliation{ Department of Physics, University of
Guilan, 41335-1914, Rasht, Iran}
\date{\today}

\begin{abstract}
 We consider isotropic spin-1/2 two-leg ladders with dominant spatially-modulated rung exchanges.
 We study the effect  of a uniform magnetic field on the ground state phase diagram of the
 model using perturbation theory and the numerical Lanczos method. The ground state phase diagram consists of two gapless Luttinger liquid (LL) and three gapped phases.  Numerically, we calculate the
 concurrence between two spins and the entanglement entropy between legs. Numerical experiment shows that the gapless LL phases are fundamentally different. In the first LL phase, only spins on rungs are entangled, but in the second LL phase the spins on legs are long-distance entangled. Therefore, the concurrence between spins on legs can be considered as a function to distinguish the LL phases.

\end{abstract}

\pacs{75.10.Jm; 75.10.Pq}

\maketitle


\section{Introduction}\label{sec1}

Entanglement is a quantum physical property that describes a
correlation between quantum systems that has no classical analogy.
When a quantum system is entangled, the best possible knowledge of
the whole does not include the best possible knowledge of its
parts. In particular, a quantum system possesses additional
correlations that do not have a classical counterpart. The study
of the nature of these quantum correlations has attracted a lot of
interest in the last decade.

At zero temperature, quantum fluctuations play the dominant
role and with the ground state properties the physics of a system determined . The quantum
fluctuations cause a fundamental change in the state of a system
which is known as the quantum phase transition. Entanglement
is expected to play an essential role in quantum phase
transitions, where quantum fluctuations manifest themselves at all
length scales. It is believed that the entanglement estimators give
important insight into the physics of the quantum phase transitions.

Low-dimensional quantum spin systems exhibit quantum phase
transitions for different driving parameters, like a magnetic
field. In particular, we address this problem by selecting a
system of spins which are arranged on a two-leg ladder
lattice with antiferromagnetic isotropic couplings along the legs
and rungs. The $S=\frac{1}{2}$ isotropic antiferromagnetic two-leg
ladders, have attracted much interest for a number of reasons.
On the one hand, there has been remarkable progress in recent
years in the fabrication of such ladder compounds\cite{dagotto99}.
On the other hand, spin-ladder systems pose interesting theoretical
problems, since they have a gap in their excitation spectrum. In the presence of a magnetic
field, they reveal an extremely rich behaviour, dominated by
quantum effects. These quantum phase transitions have been
investigated intensively both theoretically\cite{chitra97,
cabra97, mila98, totsuka98, usami98, giamarchi99, hagiwara00,
wang00, hikihara01, wessel01, wang02, wang03, hida04, vekua04} and
experimentally\cite{chaboussant97, chaboussant198, chaboussant298,
arcon99, mayaffre00, watson01}.

In a recent paper, the ground state phase diagram of a different kind of two-leg ladder with
spatially-modulated spin-exchange coupling constants has been
studied\cite{Japaridze06} in the limit of strong rung exchange
and magnetic field. The Hamiltonian of the model is defined as
\begin{eqnarray}
{\cal H} &=& J_{\parallel} \sum_{n,\alpha} {\bS}_{n,\alpha} \cdot
{\bS}_{n+1,\alpha} - H  \sum_{n,\alpha}
S^{z}_{n,\alpha} \nonumber\\
&+& J_{\perp} \sum_{n} \left[1 + (-1)^{n} \delta \right]
{\bS}_{n,1} \cdot {\bS}_{n,2},  \label{Hamiltonian}
\end{eqnarray}
where $\bS_{n,\alpha}$ is the spin $S=1/2$ operator on rung n
(n=1,...,L) and leg $\alpha$ ($\alpha=1,2$). The interleg coupling
is antiferromagnetic, $J^{\pm}_{\perp} = J_{\perp}(1 \pm \delta)
> 0 $. The non-modulated case\cite{vekua03, Mahdavifar07}, $\delta=0$, undergos two quantum phase transitions at critical fields, $H_{c_{1}}$ and $H_{c_{2}}$. The first corresponds to the
transition from the gapped rung-singlet phase to the gapless LL  phase. The second represents the transition
from the LL phase into the fully polarised ferromagnetic phase. Recently,  the bipartite entanglement measures close to the quantum critical points\cite{Tribedi09} have been investigated. It was found that the derivatives of the entanglement measures diverge as $H\longrightarrow H_{c_{2}}$ but remain finite as $H\longrightarrow H_{c_{1}}$.  The entanglement entropy was also studied for two-leg ladders within spin wave theory and the DMRG method \cite{Song11}.

For the modulated case, $\delta\neq0$, the continuum-limit
bosonizatin analysis\cite{Japaridze06} shows two additional
quantum phase transitions between a middle gapped phase and two gapless LL phases. These transitions manifest themselves most clearly in the presence of a new magnetisation plateau at magnetisation equal to
one half of its saturation value. Using the numerical Lanczos
method, the existence of four critical fields, the
magnetisation plateau at half of the saturation value and two gapless LL phases are
confirmed\cite{Mahdavifar09}.

In this paper we continue the study of this model. First, using perturbation theory, we find four critical fields in  good agreement with the previous results. Then, using the numerical Lanczos method, the concurrence and entanglement entropy will be calculated and used to show that the LL phases are fundamentally different.

The outline of the paper is as follows. In the forthcoming section we present the perturbation results. In section III we present the results of an accurate numerical experiment on the entanglement between spins at zero temperature. Finally, we discuss and summarise our results in section IV.


\section{Perturbation results} \label{sec2}

We are interested in the behaviour of the model
Eq.(\ref{Hamiltonian}) in the limit of strong rung exchange,
$J_{\perp}, H \gg J_{\parallel}$. For this aim, it is convenient to
rewrite the Hamiltonian given in Eq.(\ref{Hamiltonian}) in the form

\begin{eqnarray}
{\cal H} &=& {\cal H}_{0}+{\cal V},\nonumber \\
\nonumber \\
{\cal H}_{0} &=& J_{\perp} \sum_{n} \left[1 + (-1)^{n} \delta
\right] {\bS}_{n,1} \cdot {\bS}_{n,2}\nonumber \\
&-& H \sum_{n,\alpha} S^{z}_{n,\alpha},\nonumber \\
\nonumber \\
{\cal V}&=&J_{\parallel} \sum_{n,\alpha} {\bS}_{n,\alpha} \cdot
{\bS}_{n+1,\alpha}.  \label{Per-Hamiltonian}
\end{eqnarray}
The unperturbed part, ${\cal H}_{0}$, is the Hamiltonian of a non-interacting $N/2$
 pair of spins. The eigenstate of the unperturbed Hamiltonian is written as a product of pair states.
 By solving the eigenvalue equation of
an individual pair, one can easily find the eigenstates as
$|S\rangle=\frac{1}{\sqrt{2}}(|\downarrow\uparrow\rangle-|\uparrow\downarrow\rangle)$,
$|T_{1}\rangle=|\uparrow\uparrow\rangle$,
$|T_{0}\rangle=\frac{1}{\sqrt{2}}(|\downarrow\uparrow\rangle+|\uparrow\downarrow\rangle)$,
$|T_{-1}\rangle=|\downarrow\downarrow\rangle$. Their eigenvalues
are, respectively, $E_{|S\rangle}=-\frac{3}{4}J_{\perp}(n)$,
$E_{|T_{1}\rangle}=\frac{1}{4}J_{\perp}(n)-H$ ,
$E_{|T_{0}\rangle}=\frac{1}{4}J_{\perp}(n)$,
$E_{|T_{-1}\rangle}=\frac{1}{4}J_{\perp}(n)+H$. When the magnetic
field is zero, the ground state of a pair of spins is singlet. By
increasing the magnetic field, the energy of $|T_{1}\rangle$ becomes
closer to $E_{|S\rangle}$ and an abrupt quantum phase
transition\cite{mila98} occurs at $H_{c}=J_{\perp}(n)$. By further increasing the field, $|T_{1}\rangle$ will become the ground state of a
distinct pair. Thus, for a strong enough magnetic field, we
have a situation in which the behaviour of the system is determined by the singlet $|S\rangle$  and triplet
$|T_1\rangle$ states.

Let us start with the case of $\delta=0$. Since the ground state
energy of a distinct pair is twofold degenerate at $H_{c} =
J_{\perp}$, the ground state energy of the unperturbed Hamiltonian, $H_{0}$,
is $2^{L}$ times degenerate \cite{mila98}. The
perturbation ${\cal V}$ splits this degeneracy. By applying first order perturbation theory for finite ladders ($L =N/2=
2,3,4,5,6$) with periodic boundary conditions and generalising
the results to the thermodynamic limit we have found the critical
fields to be

\begin{eqnarray}
H_{c_{1}} &=& J_{\perp} - J_{\parallel}, \nonumber \\
H_{c_{2}} &=& J_{\perp} + 2J_{\parallel}, \label{Critical-zero}
\end{eqnarray}
which are in good agreement with the results obtained using analytical field theory\cite{vekua03, Tribedi09}. It was shown that the first
critical field, $H_{c_{1}}$, corresponds to the transition from
the gapped rung singlet phase to the LL phase. The
second one, $H_{c_{2}}$, represents the transition from the
LL phase to the fully polarised ferromagnetic
phase\cite{vekua03, Mahdavifar07}.

In the case of dimerised ladders\cite{Japaridze06, Mahdavifar09}, $\delta\neq 0$, there are two
kind of rungs: strong and weak. In this case, by increasing the
magnetic field, first weak rungs start to melt and undergo into the triplet state
with respect to the strong rungs. Therefore, it is natural to find
two additional critical fields in compare the non-dimerised
case ($\delta=0$). However, we found that second
order perturbation theory should be used for solving the eigenvalue
equation. The obtained critical fields are

\begin{eqnarray}
H_{c_{1}}&=&(1-\delta)J_{\perp} - \frac{J_{\parallel}^{2}}{2\delta
J_{\perp}},\nonumber \\
H_{c^{-}}&=&(1-\delta)J_{\perp},\nonumber \\
H_{c^{+}}&=&(1+\delta)J_{\perp} + J_{\parallel},\nonumber \\
H_{c_{2}}&=&(1+\delta)J_{\perp} + J_{\parallel} +
\frac{J_{\parallel}^{2}}{2\delta J_{\perp}}. \label{hc1af dimer}
\end{eqnarray}
We emphasise that the above perturbation results are in good agreement with estimations made within the continuum-limit
approach\cite{Japaridze06} and numerical Lanczos
method\cite{Mahdavifar09}. It has been shown\cite{Japaridze06, Mahdavifar09} that the ground state phase diagram consists of five phases: (I.) the rung-singlet phase in the region $H<H_{c_{1}}$, (II.) the first LL phase in the region $H_{c_{1}}<H<H_{c^{-}}$, (III.) the gapped mid-plateau state in the region $H_{c^{-}}<H<H_{c^{+}}$, (IV.) the second LL phase in the region $H_{c^{+}}<H<H_{c_{2}}$ and (V.) the saturated ferromagnetic phase in the region $H>H_{c_{2}}$.

On the other hand, we also calculated the ground state of
the system in different sectors of the ground state phase diagram, which allows us to follow the melting process on the entanglement
function\cite{Wooters98, Amico08}. The entanglement of formation is defined as
\begin{eqnarray}
 E = -Xlog_{2}X-(1-X)log_{2}(1-X), \label{formation}
\end{eqnarray}
where $X=\frac{1}{2}(1+\sqrt{1-C^{2}})$ and $C$ is the concurrence which is given by
\begin{eqnarray}
C_{lm}&=&2~max\{0, C_{lm}^{(1)}, C_{lm}^{(2)}\},\nonumber
\label{concurrence}
\end{eqnarray}
where
\begin{eqnarray}
C_{lm}^{(1)}&=& \sqrt{(g_{lm}^{xx}-g_{lm}^{yy})^{2}+(g_{lm}^{xy}+g_{lm}^{yx})^{2}} \nonumber \\
&-&\sqrt{(\frac{1}{4}-g_{lm}^{zz})^{2}-(\frac{M_{l}^{z}-M_{m}^{z}}{2})^{2}},\nonumber \\
C_{lm}^{(2)}&=& \sqrt{(g_{lm}^{xx}+g_{lm}^{yy})^{2}+(g_{lm}^{xy}-g_{lm}^{yx})^{2}} \nonumber \\
&-&\sqrt{(\frac{1}{4}+g_{lm}^{zz})^{2}-(\frac{M_{l}^{z}+M_{m}^{z}}{2})^{2}}.
\label{concurrence}
\end{eqnarray}
$g_{lm}^{\alpha\beta}=\langle S_{l}^{\alpha} S_{m}^{\beta}\rangle$
denotes the correlation function between spins on sites $l$ and
$m$ and $M_{l}^{z}=\langle S_{l}^{z} \rangle$. The notation $\langle ... \rangle$ represents the ground
state expectation value. Because $J_{\perp} \gg J_{\parallel}$, the entanglement of formation
between spins on a rung is an important quantity. In the region of the
magnetic filed, $H<H_{c_{1}}$, the ground state is written as a
product of singlet rung states. In this region the concurrence
between spins on a rung attains its maximum value, one. For very large
values of the field, $H>H_{c_{2}}$, the system is in the saturated
ferromagnetic phase and the ground state is a product of triplet
($|T_{1}\rangle$) rung states. In the saturated phase the
concurrence takes zero value. Therefore, during the melting
process, the concurrence between spins on rungs changes from the completely entangled state into the non-entangled state.
In addition, using the perturbation results on finite ladders, we found a
relation for the concurrence between spins on a rung in different subspaces versus the number of rungs as
\begin{eqnarray}
C = 1-\frac{i}{L},\label{C}
\end{eqnarray}
where $i$ is the number of triplet rungs in a subspace. In the rung singlet phase, the number of triplet rungs is zero and the concurrence is maximum. On the other hand, in the saturated ferromagnetic phase all rungs are triplet, $i=L$,  and particles are not entangled ($C=0$).


\section{Numerical experiment} \label{sec3}

By performing an experiment, one can obtain a clear picture of the entanglement
phenomena in the ground state magnetic phases of the model. Since a real experiment cannot be done at zero
temperature, the best way is to undertake a virtual numerical experiment. A very famous and accurate method in field of numerical experiments is known as the Lanczos method\cite{Lanczos50,Grosso95}. In this section, to explore the nature of the spectrum and the
quantum phase transition, we used the Lanczos method to numerically diagonalise
 ladders with lengths up to $N=2L=28$ and different
values of the exchanges\cite{Mahdavifar09}. The energies of the
lowest-energy eigenstates were obtained for ladders with periodic
boundary conditions.

\subsection{CRITICAL FIELDS}

Since in a critical field the energy gap should be closed, the
best way to find the critical fields is to investigate the
energy gap, which is recognised as the difference between the energies of the first exited state and
the ground state in finite two-leg ladders. As previous
section, we start our consideration with the case of isotropic and
uniform rung exchange ladders ($\delta=0$). We have calculated the energy gap for ladders with
exchanges $J_{\parallel}=1$, $J_{\perp}=6$ and lengths
$N=12, 16, 20, 24, 28$. By means of the phenomenological renormalisation
group technique\cite{Mahdavifar08-1} we have found
$H_{c1}=5.04\pm0.01$ and $H_{c_{2}}=8.12\pm 0.01$, in good agreement
with the perturbation results (Eq.(\ref{Critical-zero})).

We have also considered the case of a ladder with alternating rungs
$\delta\neq0$. Critical fields calculated numerically for
ladders with exchange parameters $J_{\perp}=\frac{11}{2}$,
$\delta=\frac{1}{11}$, $J_{\parallel}=1$ are\cite{Mahdavifar09}

\begin{eqnarray}
H_{c_{1}}&=&4.47\pm0.01,\nonumber \\
H_{c^{-}}&=&5.34\pm0.01,\nonumber \\
H_{c^{+}}&=&6.87\pm0.01,\nonumber \\
H_{c_{2}}&=&7.62\pm0.01,  \label{critical point2}
\end{eqnarray}
which are in good agreement with the second-order perturbation
results (Eq.(\ref{hc1af dimer})).

\begin{figure}[t]
\centerline{\psfig{file=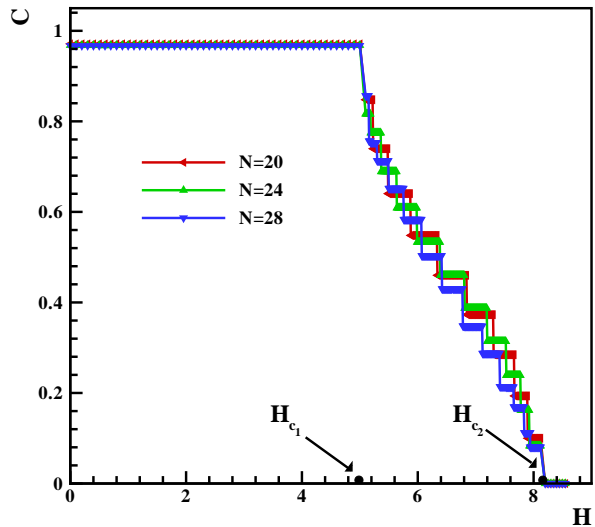,width=3.65in}}
\caption{(colour online). The concurrence between spins on rungs as a
function of the external magnetic field for ladders with lengths
$N=20, 24, 28$ and uniform rung couplings, $\delta=0$, $J_{\perp}=6.0$ and $J_{\parallel}=1.0$.}
\label{concurrence-uniform}
\end{figure}

\subsection{CONCURRENCE}

In this section we focus on the entanglement of formation as a
measure of the entanglement. We compute the entanglement between
two spins, which is known as the concurrence
(Eq.(\ref{concurrence})). In the region
$H<H_{c_{1}}$, where the system is in the non-magnetic
rung-singlet phase, the correlations $g^{xx}$, $g^{yy}$, $g^{zz}$
should be the same and take the value $-1/4$, and therefore the concurrence
must take the maximum value $C=1$. On the other hand, for very large values of the
magnetic field, $H>H_{c_{2}}$, the ground state of the system is
written as
\begin{eqnarray}
|Gs\rangle=| \uparrow\uparrow\uparrow...\uparrow\rangle.
\label{dimer-GS}
\end{eqnarray}
In this saturated ferromagnetic phase, the correlations are $g^{xx}=g^{yy}=0, g^{zz}=1/4$ and $M^{z}=1/2$. Therefore, the concurrence takes the minimum value $C=0$.

The numerical Lanczos results describing the concurrence for a two-leg
ladder with antiferromagnetic legs and the same rungs ($\delta=0$)
are shown in Fig.~\ref{concurrence-uniform}. In this figure the
concurrence between two spins on a rung is plotted as a
function of $H$ for ladder lengths $N=20, 24, 28$  and
exchanges $J_{\parallel}=1$, $J_{\perp}=6$. It can be seen that the numerical results are
in reasonable agreement with what is expected. Essentially our
numerical experiment shows that in the absence of a magnetic
field, spins on a rung are maximally entangled. By applying a
magnetic filed up to the first critical field $H_{c_{1}}$, the
concurrence remains constant. This behaviour is in agreement with
expectations, based on the general statement that in the gapped
rung-singlet phase, the change in any physical function appears
only at a finite critical value of the magnetic field equal to the
spin gap. The concurrence starts to decrease as soon as the field
passes beyond  the first critical field $H_{c_{1}}$. Indeed the quantum correlations of the two spins on rungs decrease with increasing magnetic field.  In the
intermediate region the decreasing behaviour of the concurrence
continues up to the second critical field $H_{c_{2}}$, where it takes
the zero value. As expected in the region of the saturated
ferromagnetic phase, $H>H_{c_{2}}$, the concurrence remains zero. The oscillations of the concurrence at finite $N$ in the region $H_{c_{1}}<H<H_{c_{2}}$, are the result of level
crossings between the ground state and excited states of
the model.

\begin{figure}[t]
\centerline{\psfig{file=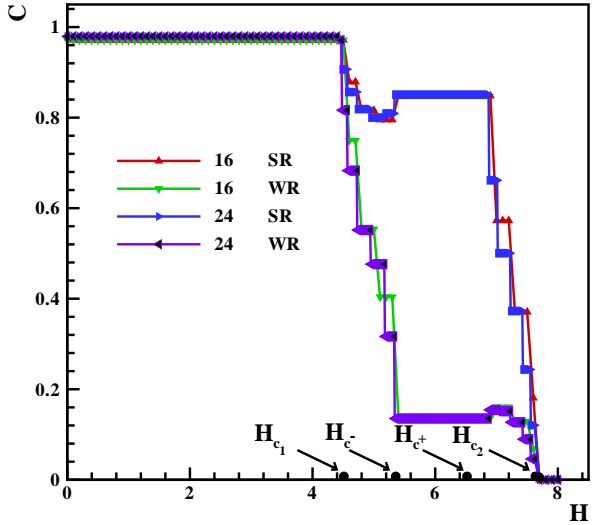,width=3.65in}}
\caption{(colour online). The concurrence between spins on strong (SR) and weak (WR) rungs versus applied magnetic
field for dimerised ladders with
lengths $N=16, 24$ and exchange parameters
$\delta=\frac{1}{11}$, $J_{\perp}=\frac{11}{2}$ and
$J_{\parallel}=1.0$. }
\label{concurrence-dimerized}
\end{figure}

Let us now present our numerical results for the dimerised case
$\delta\neq 0$. In Fig.~\ref{concurrence-dimerized} we have plotted
the concurrence of two spins on strong $(J_{\bot}(1+\delta))$ and weak $(J_{\bot}(1-\delta))$ rungs as a function of the
magnetic field $H$. We have considered two-leg ladders with
different lengths $N=16, 24$ and exchange parameters
$J_{\perp}=\frac{11}{2}$, $\delta=\frac{1}{11}$,
$J_{\parallel}=1$. It can be seen that spins on weak and strong rungs are maximally entangled at $H=0$.  By increasing the magnetic field from zero, the spins on strong and weak rungs remain maximally entangled up to the first critical field $H=H_{c_{1}}$.  For $H>H_{c_1}$ the concurrence between spins on weak rungs drops very rapidly when compared to the concurrence between spins on strong rungs. Indeed, the quantum correlations of two spins on  strong rungs and weak rungs decrease with increasing magnetic field, but with the different intensity. In the intermediate gapless LL region $H_{c_1}<H<H_{c}^{-}$, the quantum correlations of strong and weak rungs diminish down to $H=H_{c}^{-}$ and the concurrences reduce to $\sim 0.8$ and $\sim 0.15$ respectively.  In the half-plateau state, the gap of the system is re-opened and a plateau emerges in the curve of concurrences with the values $\sim 0.85$ and $\sim 0.15$. By further increasing the magnetic field and for $H>H_{c}^{+}$, the concurrence between spins on strong rungs drops very rapidly when compared to the concurrence on weak rungs. Finally, in the full-saturated ferromagnetic state, all of concurrences disappear and the entanglement of the state is exactly zero.

\begin{figure}[t]
\centerline{\psfig{file=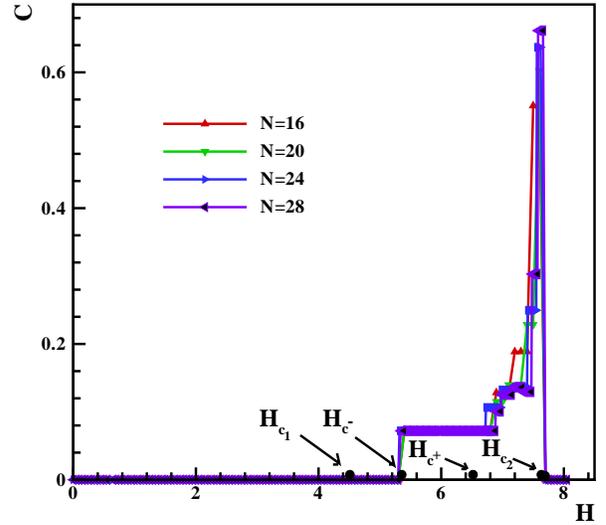,width=3.65in}}
\caption{(colour online). The concurrence between nearest neighbour (NN) spins on legs as a
function of the magnetic field $H$ for dimerised ladders with lengths $N=16, 20, 24, 28$
and exchange parameters $\delta=\frac{1}{11}$,
$J_{\perp}=\frac{11}{2}$ and $J_{\parallel}=1.0$.}
\label{antiferro.nn}
\end{figure}

Apart from the concurrence between spins on rungs, it is completely natural to ask about the concurrence between spins on legs. To find a clear answer, we have calculated the concurrence between spins on the legs.  We have plotted the concurrence between two spins which are nearest neighbours (NN) on a leg in Fig.~\ref{antiferro.nn}. The numerical results in this figure are computed for  two-leg ladders with
different lengths $N=16, 20, 24, 28$ and exchange parameters
$J_{\perp}=\frac{11}{2}$, $\delta=\frac{1}{11}$,
$J_{\parallel}=1$. It is clearly seen that the NN spins on legs are not entangled in the absence of a magnetic filed. By applying a magnetic field, the concurrence between NN spins remains zero up to the critical field $H=H_{c^{-}}$. By increasing the magnetic field beyond this value, the NN spins on legs become entangled in the plateau state $H_{c^{-}}<H<H_{c^{+}}$.  This means that the magnetic field increases the quantum correlations of the two spins which are NN on a leg. This is a dual effect of the magnetic field, in which increases in the quantum correlations of two NN spins on legs accompany decreases in the quantum correlations of spins on rungs. Indeed, in the plateau state there are three types of quantum correlations in the system.
These correlators are the source of the mid-plateaus in the different parameters of the system, such as magnetisation. with further increases in the magnetic field $H>H_{c^{+}}$, the concurrence between NN spins decreases and becomes zero at the saturation magnetic field $H=H_{c_{2}}$.

\begin{figure}[t]
\centerline{\psfig{file=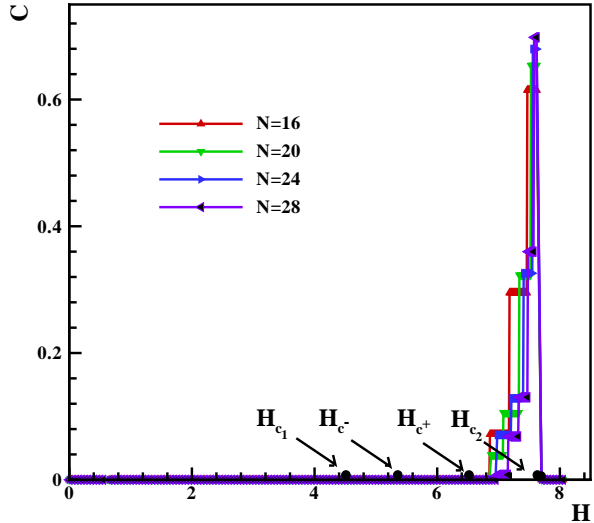,width=3.65in}}
\caption{(colour online). The concurrence between next nearest neighbour (NNN) spins on legs as a
function of the magnetic field $H$ for dimerised ladders with lengths $N=16, 20, 24, 28$
and exchange parameters $\delta=\frac{1}{11}$,
$J_{\perp}=\frac{11}{2}$ and $J_{\parallel}=1.0$.}
\label{antiferro.nnn}
\end{figure}

\begin{figure}[t]
\centerline{\psfig{file=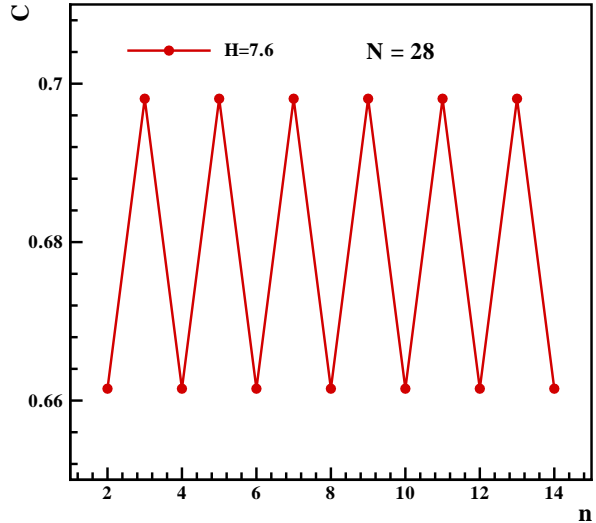,width=3.65in}}
\caption{(colour online). The concurrence between spins on legs in the second LL phase versus the separation distance $n$ for a ladder with length $N=28$ and exchanges $\delta=\frac{1}{11}$,
$J_{\perp}=\frac{11}{2}$ and $J_{\parallel}=1.0$.  } \label{long.distance}
\end{figure}

To get more information about the entanglement between spins, we have also calculated the concurrence between next nearest neighbours (NNN) on legs. The numerical results on the NNN spins on a leg are plotted in Fig.~\ref{antiferro.nnn} for   two-leg ladders with
lengths $N=16, 20, 24, 28$ and exchange parameters
$J_{\perp}=\frac{11}{2}$, $\delta=\frac{1}{11}$,
$J_{\parallel}=1$. It is completely clear that the NNN spins are not entangled at $H=0$. In the presence of a magnetic field the NNN spins will be entangled only in the second gapless LL phase, $H_{c^{+}}<H<H_{c_{2}}$. This is related to the fact that in the second LL phase, the magnetic field increases the quantum correlations of the two spins which are NNN spins on a leg. The concurrence between them is equal to zero in the mid-plateau state. Our numerical results show that only NN spins on legs are entangled in the mid-plateau state. As in the case of NN spins, the concurrence between NNN spins decreases and becomes zero at the saturation magnetic field $H=H_{c_{2}}$. To gain further insight into the second LL state, we have computed numerically the entanglement between the spins with different
separation distances for ladders with lengths up to $N=28$. The concurrence between an arbitrary spin and another spin, say $S_n$ has been  plotted versus the separation distance, $n$, in Fig.~\ref{long.distance}. It is very interesting that the entanglement of
two such spins on legs does not vanishes by increasing the distance between them. It means that, in the second LL state, $H_{c^{+}}<H<H_{c_{2}}$, the range of the quantum
correlations between spins on the legs is long-distance. There is a non-vanishing long-distance entanglement
between two spins on legs in dimerised ladder systems. The oscillation with period two is related to the fact that in the dimerised ladders there are  two kind  of rungs.



\subsection{ENTANGLEMENT ENTROPY} \label{sec2}

Up to this point we have presented the results of the numerical calculations on the
entanglement of formation, which is useful for studying the entanglement between a pair of particles. However in some cases, two  groups of particles can be entangled. Measurements of entanglement between subsystems, chiefly
using entropic quantities, have an advantage over traditional
two-point correlation functions in that they encode the total
amount of information shared between the subsystems without the
possibility of missing hidden correlations.

The entanglement entropy (EE) is defined as the von Neumann
entropy of a reduced density matrix of a subsystem. This function
is useful for understanding the different quantum phases in condensed
matter systems. The EE is defined as
\begin{eqnarray}
E^{vN}=-\langle \log \hat{\rho}_{A} \rangle=-Tr_{A}[\hat{\rho}_{A}\log
\hat{\rho}_{A}], \label{EE-formula}
\end{eqnarray}
where, $\hat{\rho}_{A}=Tr_{B}[\hat{\rho}]$ and $\hat{\rho}$ is the
density matrix of the ground state $|Gs \rangle$. It is assumed
that the system consists of subsystems $A$ and $B$. The EE
quantifies the information describing how much the state is entangled
between the subsystems $A$ and $B$.

\begin{figure}[t]
\centerline{\psfig{file=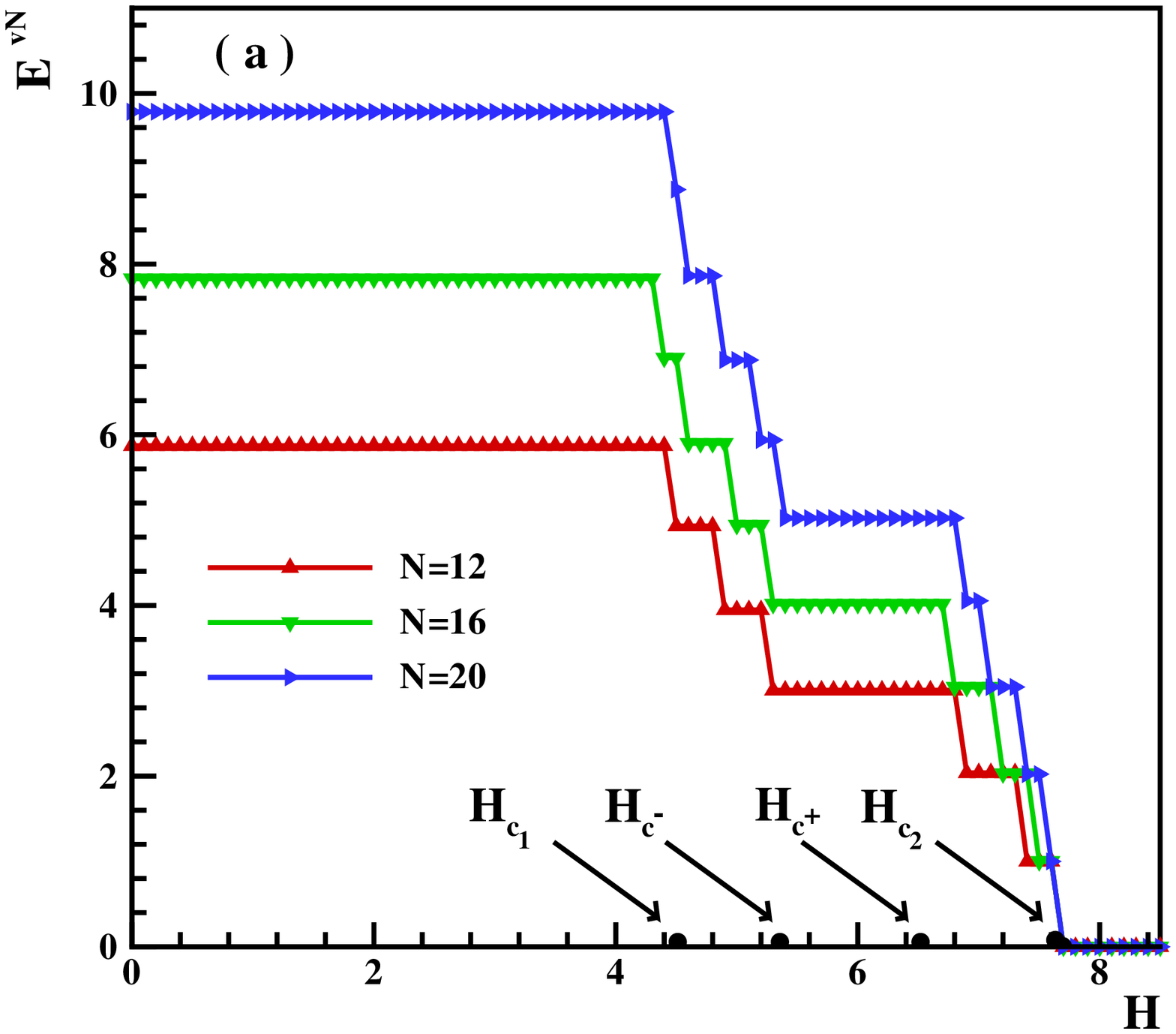,width=3.65in}}
\centerline{\psfig{file=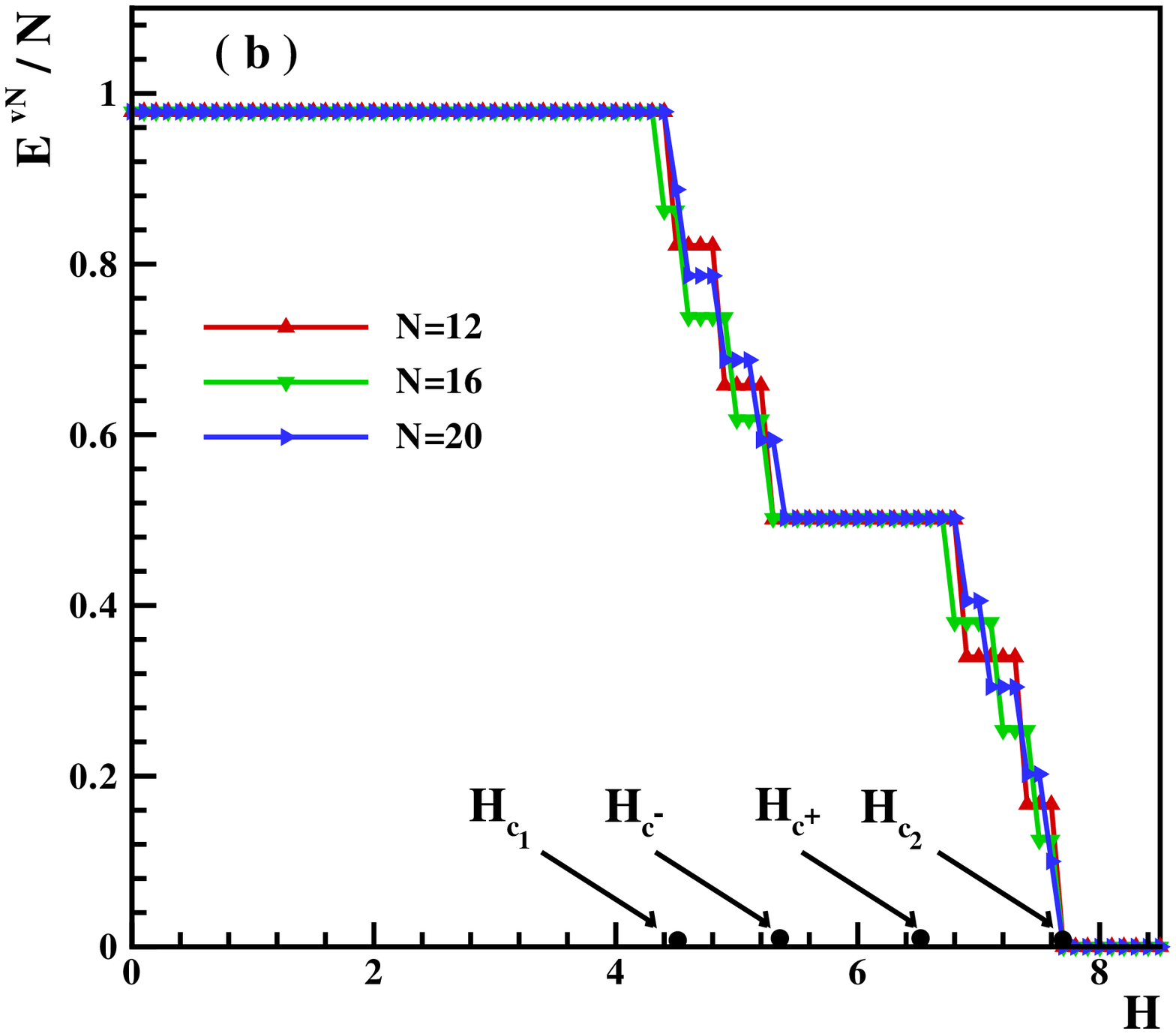,width=3.65in}}
\caption{(colour online.) (a) The entanglement entropy between two
legs for  dimerised ladder with lengths $N=12,16,20$ and exchanges $\delta=\frac{1}{11}$,
$J_{\perp}=\frac{11}{2}$ and $J_{\parallel}=1.0$. (b) The entanglement entropy per rungs as a function of the magnetic field $H$. } \label{von Neumann}
\end{figure}

In our numerical experiment, legs are selected as subsystems A and B. The Lanczos  numerical results on the function $E^{vN}$, are plotted in Fig.~\ref{von Neumann}(a) for two-leg ladders with
lengths $N=12, 16, 20$ and exchange parameters
$J_{\perp}=\frac{11}{2}$, $\delta=\frac{1}{11}$,
$J_{\parallel}=1$. It is Clearly seen that the legs are entangled in the absence of an external magnetic field. They remain entangled in the presence of a magnetic field up to the first critical field $H_{c_{1}}$. It has been found that, away from the critical points, the EE scales with the size of the boundary between A and B, which is known as the area law\cite{Srednicki93}. It is completely clear that in the rung-singlet phase, $H<H_{c_{1}}$, the EE is proportional to the number of rungs, which is known as the area law in ladder sytems\cite{Kallin09, Poilblanc10}. On the other hand,  a different concept of entanglement has been suggested, known as valence bond entanglement entropy\cite{Capponi07, Chhajlany07}, $E^{VB}=\ln(2).p$, where $p$ is the number of singlets crossing the boundary between regions $A$ and $B$. It is also clearly seen that in the singlet region, $H<H_{c_{1}}$, the valence bond entropy is less than the von Neumann entropy, in good agreement with Ref.[\cite{Kallin09}]. As soon as the magnetic field increases from the first critical field, $H>H_{c_{1}}$, the entanglement entropy decreases and the process stops at the critical field $H_{c^{-}}$.  Like the other physical functions, a plateau appears in curve of the EE function in the intermediate gapped state, $H_{c^{-}}<H<H_{c^{+}}$. Finally, by further increasing the magnetic field, $H>H_{c^{+}}$, the EE decreases and vanishes at the saturation magnetic field $H_{c_{2}}$. The area law suggests that one can find a better picture by dividing the entanglement entropy, $E^{vN}$, by the number of rungs, $L$.  In Fig.~\ref{von Neumann}(b) we have therefore  plotted $E^{vN}/L$ as a function of the magnetic field $H$.  It is clear that legs are maximally entangled in the region $H\leq H_{c_{1}}$, and that in the intermediate gapped plateau region, $H_{c^{-}}<H<H_{c^{+}}$, the entanglement entropy is almost equal to half of the saturation value. There is clearly a decreasing process in both the LL regions $H_{c_{1}}\leq H\leq H_{c-}$ and $H_{c+}\leq H\leq
H_{c_{2}}$.


\section{conclusion}\label{sec-III }

In this paper, we have focused on the entanglement between spins in isotropic spin-1/2 two-leg ladders with dominant spatially-modulated rung exchanges. Using perturbation theory and the numerical Lanczos method we have studied the effects of an external magnetic field and a space modulation on the concurrence and the entanglement entropy. By employing perturbation theory, four critical fields in good agreement with the previous Bosonization results are found. Using the perturbation results on finite ladders, we have found a
relationship  between the concurrence between spins on a rung and the number of rungs (Eq.\ref{C}).

Moreover, we implemented the Lanczos method to numerically
diagonalise ladders with finite length up to $N=28$. Using the
exact diagonalisation technique,  we calculated the energy gap,
concurrence and the entanglement entropy between legs for different values of the external
magnetic field. In the intermediate gapped state, we found that a non-zero plateau also appears in the plot of concurrence between spins on rungs and nearest neighbour spins on legs. On the other hand, our numerical results on the concurrence showed that the two gapless LL phases in the ground state phase diagram of the model  are fundamentally different. In one of them only spins on rungs are entangled, but in the next LL phase the spins on legs are long-distance entangled. Thus the concurrence between spins on legs can be considered as a function to distinguish the LL phases.

\section{acknowledgments}
It is our pleasure to thank G. I. Japaridze for very useful comments and interesting discussions. We are also grateful to M. Motamedifar for carefully reading our manuscript and appreciate
his useful comments.




\vspace{0.3cm}

\end{document}